\documentclass[a4paper,11pt]{article}
\usepackage{jheppub} % for details on the use of the package, please
\usepackage{natbib}
\usepackage{tensor}

\usepackage{physics,amsmath,amsfonts,amsthm,amssymb, bm, mathtools, latexsym}

\renewcommand{\a}{\alpha}

\newcommand{\pd}{\partial}

\renewcommand{\d}{\delta}

\newcommand{\imp}{\implies}

\newcommand{\D}{\Delta}
\newcommand{\e}{\ep}
\newcommand{\ep}{\varepsilon}

\title{\boldmath Rindler Fluids from Gravitational Shockwaves}

\author[a]{Sang-Eon Bak,}
\author[a]{Cynthia Keeler,}
\author[b]{Yiwen Zhang,}
\author[b]{and Kathryn M. Zurek}

\affiliation[a]{Department of Physics, Arizona State University, Tempe, AZ 85281, USA}
\affiliation[b]{Walter Burke Institute for Theoretical Physics, Caltech, Pasadena, CA 91125, USA}
\emailAdd{sbak2@asu.edu}
\emailAdd{keelerc@asu.edu}
\emailAdd{yiwenz@caltech.edu}
\emailAdd{kzurek@caltech.edu}

\abstract{We study a correspondence between gravitational shockwave geometry and its fluid description near a Rindler horizon in Minkowski spacetime. Utilizing the Petrov classification that describes algebraic symmetries for Lorentzian spaces, we establish an explicit mapping between a potential fluid and the shockwave metric perturbation, where the Einstein equation for the shockwave geometry is equivalent to the incompressibility condition of the fluid, augmented by a shockwave source.   Then we consider an Ansatz of a stochastic quantum source for the potential fluid, which has the physical interpretation of shockwaves created by vacuum energy fluctuations. Under such circumstance, the Einstein equation, or equivalently, the incompressibility condition for the fluid, becomes a stochastic differential equation. 
By smearing the quantum source on a stretched horizon in a Lorentz invariant manner with a Planckian width (similarly to the membrane paradigm), we integrate fluctuations near the Rindler horizon to find an accumulated effect of the variance in the round-trip time of a photon traversing the horizon of a causal diamond.
}

\begin{document} 
\maketitle
\flushbottom

\section{Introduction}
\label{sec:intro}

Almost fifty years ago, Damour~\cite{Damour3598} first connected general relativity to fluid dynamics by demonstrating that when perturbations fall into a black hole horizon, the spacetime near the horizon behaves like a fluid. This idea led to the conception of the membrane paradigm~\cite{thorne1986black, Parikh:1997ma,Eling:2009sj,Eling:2009pb, Gourgoulhon:2005ch}, in which the fluid resides on a stretched horizon, a timelike hypersurface very close to the actual black hole horizon. Later on, the discovery of the AdS/CFT correspondence formalized a version of fluid/gravity duality in which the fluid arises as an effective description in the long distance, low frequency regime of the dual gauge theory on the AdS boundary~\cite{Policastro:2001yc,Policastro:2002se,Kovtun:2003wp,Kovtun:2004de,Son:2007vk,Bhattacharyya:2008kq,Faulkner:2010jy}. 

More recently, substantial progress was made in studying fluid/gravity duality on a cutoff surface in flat Rindler spacetime~\cite{Bredberg:2010ky,Bredberg:2011jq,Compere:2011dx,Compere:2012mt,Lysov:2011xx,Pinzani-Fokeeva:2014cka}. The flat space cutoff approach to fluid/gravity duality formulates the map between the Einstein and Navier-Stokes equations in a precise manner. In particular, Refs.~\cite{Bredberg:2011jq,Compere:2011dx,Pinzani-Fokeeva:2014cka}  showed that the fluid is defined in terms of the extrinsic curvature of a cutoff hypersurface outside of the horizon. This cutoff surface is equipped with a flat induced metric. As in~\cite{Bredberg:2011jq}, in the leading order hydrodynamic expansion, the constraint equation from the Einstein tensor becomes the incompressibility condition
\begin{equation} \label{eq:intro incomp}
	\pd_i v^i = 0.
\end{equation} 
Here $v^i$ is the $i$-component of the fluid velocity. Furthermore, the next leading order constraint Einstein equation becomes the incompressible Navier-Stokes equation~\cite{Bredberg:2011jq}
\begin{equation} \label{eq:intro Navier stokes}
	\pd_{\tau} v_i - r_c \pd^2 v_i + \pd_i P +v^j \pd_j v_i = 0.
\end{equation} 
Thus the cutoff surface fluid-gravity duality relates Einstein gravity, which governs nonlinear gravitational interactions, to
the incompressible Navier-Stokes equations, which are an effective description of classical fluids.

In this paper, we aim to study the connection between gravitational shockwave spacetimes and their fluid descriptions. Gravitational shockwave geometries were studied first by Dray and 't Hooft~\cite{DRAY1985173} when they considered ultra-relativistic matter falling into a black hole. More recently, \cite{Cristofoli:2020hnk,Gray:2021dfk,Verlinde_2022,He:2023qha} have studied shockwaves in flat spacetimes on the quantum level. Rewriting these gravitational shockwaves on the fluid side of fluid/gravity duality will allow us to consider quantum shocks as a fluid source.

Technically, we match gravitational shockwaves to their dual fluids by using their Petrov type. Since the shockwave geometries we consider are Petrov type N, their fluid dual must correspond to a spacetime of the same type.  As shown in ~\cite{Keeler:2020rcv}, potential fluids with $v_i=\nabla_i \phi$ have dual geometries of Petrov type N.
Since Petrov Type N spacetimes only allow one degree of freedom, we map between the cutoff fluid formalism and shockwave geometries by identifying the fluid potential with the shockwave metric via
\begin{equation} \label{eq:intro phi}
	\phi = \beta r H_{uu}.
\end{equation}  
Here $\phi$ is the fluid potential and $H_{uu}$ characterizes the metric fluctuations due to a shockwave. By choosing $\beta$ carefully, we find an exact mapping such that the Einstein equation for the shockwave geometries is precisely the incompressibility condition in Eq.~\eqref{eq:intro incomp} with the right hand side replaced by the shockwave stress-energy tensor. Solving this differential equation allows us to constrain the pressure term in the next leading order Navier-Stokes equation.

Since we map the shockwaves to potential fluids, the next major goal of our paper is to study the effects of quantum fluctuations on potential fluids. In the leading order hydrodynamic expansion, we keep the left hand side of the incompressibility equation unchanged, while replacing the right hand side with a quantum noise term. This Ansatz is motivated by a set of commutation relations first studied by 't Hooft~\cite{tHooft:1996rdg,tHooft:2018fxg} in the context of black hole perturbations, and later developed by~\cite{Verlinde_2022} and~\cite{He:2023qha} for flat spacetime. 
Such fluctuations alter the classical fluid (Einstein) equation to a stochastic differential equation, resembling the form of a first order Langevin equation, which describes Brownian motion of a massless particle~\cite{Zhang:2023mkf}. Quantum uncertainty arises from a Brownian particle undergoing random walk in a stochastic background~\cite{Banks:2021jwj,Zurek:2022xzl}.

We argue in a similar fashion as in~\cite{Zhang:2023mkf} that spacetime fluctuations give rise to a ``smeared out'' horizon, analogous to a stretched horizon~\cite{Damour3598}. However, our present paper utilizes the cutoff surface fluid approach to devise a Lorentz invariant method to smear out the horizon. We will impose the condition that the cutoff surface intersects with the boundary of a causal diamond at a distance of the reduced Planck length $\tilde{\ell}_p$~\cite{Banks:2021jwj,Zurek:2022xzl,Verlinde_2022}. In four dimensional spacetime, the reduced Planck length is simply the Planck length $\tilde{\ell}_p = \ell_p$. Positing such a cutoff surface allows us to compute the variance of photon traversal time measured by a free-falling observer on the boundary of the causal diamond 
\begin{equation} \label{eq:intro var T}
	\D T_{\rm r.t.}^2\sim 2 L \ell_p f(\bm{x}, \bm{x}').
\end{equation}
Here $f(\bm{x}, \bm{x}')$ is the transverse response function, which is a unique feature to the shockwave geometries, and is absent in classical Brownian motion. Eq.~\eqref{eq:intro var T} is consistent with the result from previous calculations~\cite{Verlinde:2019ade,Verlinde:2019xfb,Banks:2021jwj,Zurek:2022xzl,Zhang:2023mkf} done via different means.

The outline of this paper is as follows. In Sec.~\ref{sec:Fluid Full Section} we review the cutoff fluid formalism in flat spacetime. In Sec.~\ref{sec:shockwaves} we compute the Petrov classification for the shockwave spacetime and the fluid metric, respectively, and we show that the shockwave metric is mapped to the potential of a potential flow in hydrodynamics. In Sec.~\ref{sec:fluctuation dissipation}, we solve the incompressibility equation with a source term posited from the 't Hooft commutation relations, and use the solution to compute the uncertainty in photon traversal time. In Sec.~\ref{sec:summary}, we summarize our results and point to some future directions. Throughout this paper, we use $8 \pi G_N = \ell_p^2 = \kappa^2 $.

%%%%%%%%%%%%%%%%%%%%%%
%%%%%%%%%%%%%%%%%%%%%%
\section{Rindler Fluids in Einstein Gravity}
\label{sec:Fluid Full Section}
In this section we review the fluid/gravity correspondence on a cutoff surface of Minkowski space. Following \cite{Bredberg:2010ky, Bredberg:2011jq, Compere:2011dx, Compere:2012mt}, we develop the duality by first performing coordinate transformations to introduce a constant fluid velocity and constant pressure.  These transformations produce the ``seed'' metric, which is just Minkowski space in unusual coordinates.  Then, we allow the fluid velocity and pressure to vary slowly with both the $x^i$ and $\tau$ coordinates; allowing for higher-order corrections in the hydrodynamic parameter then produces the true fluid dual metric.  Resolving the Einstein equations for this more general metric, we find the constraint Einstein equations can be written as conservation equations for the Brown-York stress tensor.  Writing these conservations equations explicitly, the first and second-order equations become incompressibility and the Navier-Stokes equations, respectively. Following this review, we additionally review the specific case of potential fluids for later use.

\subsection{Geometric Setup}
\label{sec:fluids setup}

We will work throughout in ingoing Eddington-Finkelstein coordinates
\begin{equation} \label{eq:ingoing rindler}
	d s^2 = - (\a r) d \tau^2 + 2 d \tau d r + \delta_{i j} d x^i d x^j,
\end{equation}
where $i,j$ indicate the transverse directions and thus run from $1\ldots d-2$ (these correspond to to the 2 transverse directions in $d=4$).  %Since we will later specialize to $d=4$, $i,\,  j$ can also be thought of as running over just $1,2$.  Also, 
$\a$ corresponds to the constant proper acceleration of a Rindler observer at fixed $r$. Our formulation most closely follows \cite{Compere:2011dx,Compere:2012mt,Eling:2012ni}.

Our fluid will live on a timelike hypersurface called $\Sigma_c$.  We choose this surface to be $r = r_c$; its induced metric $\gamma_{a b}$  is
\begin{equation} \label{eq:gamma_ab}
	d s^2 |_{\Sigma_c} = \gamma_{a b} d x^a d x^b = - (\a r_c) d \tau^2 + \delta_{i j} d x^i d x^j.
\end{equation}
Here $a, b$ are spacetime indices on $\Sigma_c$, so they run over $\tau$ as well as $i,\, j$. Note that this metric is flat.

Starting from this background geometry, we introduce a constant pressure by shifting $r$; since we want to keep the induced metric fixed, we must also rescale $\tau$. The transformation is
\begin{align}\label{eq:shift}
    r \rightarrow r+\frac{1}{\alpha p^2}- r_c, \quad \tau \rightarrow \sqrt{\alpha r_c} p \tau\,.
\end{align}
This coordinate transformation shifts the position of the Rindler horizon from $r_h=0$ to  $r_h=r_c-\frac{1}{\alpha p^2}$.

We then introduce a constant fluid velocity by performing the boost
\begin{align}\label{eq:boost}
    \sqrt{\alpha r_c} \tau \rightarrow-u_a x^a, \quad x^i \rightarrow x^i-u^i \sqrt{\alpha r_c} \tau+(1+\gamma)^{-1} u^i u_j x^j\,.
\end{align}
where $\gamma= \qty(1- v^2 /(\alpha r_c))^{-1 / 2}$, and we set the fluid four-velocity vector $u_a$ to a constant. 

Thus we arrive at the `seed' metric for a relativistic fluid, given by
\begin{align}\label{eq:seed metric}
    d s^2=-2 p u_a d x^a d r+\left[\gamma_{a b}-\a p^2\left(r-r_c\right) u_a u_b\right] d x^a d x^b\,.
\end{align}
As noted above, so far this metric is just a rewriting of Minkowski space in fancy coordinates.  However, we can already begin to study the constant `fluid' on the cutoff surface.  To characterize the physics on the cutoff surface, we introduce the Brown-York stress-energy tensor~\cite{Brown:1992br}
\begin{equation} \label{eq:Tab}
	\kappa^2 T_{a b} =   (K \gamma_{a b} - K_{a b}),
\end{equation}
where $\kappa^2 =8\pi G_N$. Here, $K_{ab}= \frac{1}{2} \mathcal{L}_{n} \gamma_{a b}$ is the extrinsic curvature on $\Sigma_c$, and $K=K_{a b} \gamma^{a b}$ is its trace. Here, $\mathcal{L}_{n}$ is the Lie derivative along the unit normal vector $n^\mu$ on the cutoff surface $\Sigma_c$. 

For the seed metric in Rindler spacetime~\eqref{eq:seed metric}, the Brown-York stress tensor on the cutoff surface has a form of the perfect fluid in equilibrium with a vanishing energy density:
\begin{align}\label{eq:perfect fluid}
    \kappa^2 \,T_{a b}=\alpha p \, h_{a b}\,,
\end{align}
where $h_{a b}=\gamma_{a b}+u_a u_b$ is the metric of a spacelike hypersurface orthogonal to the velocity vector $u_a$ but still embedded in $\Sigma_c$.  Since the pressure is constant, there is no time evolution of the momentum density of the fluid. Due to the vanishing energy density, the divergence of the momentum density is vanishing as well by the continuity equation. Thus, the Brown-York stress tensor \eqref{eq:perfect fluid} for the seed metric describes the equilibrium fluid that does not flow as seen by a comoving observer with $u^a$.  Thus this background seed metric is equivalent to a somewhat boring fluid. 

Before we introduce slow variations in order to consider more interesting fluids, we note that the Brown-York stress-energy tensor $T_{ab}$ is not equivalent to the external matter stress-energy tensor $\mathcal{T}_{\mu\nu}$. The matter stress-energy tensor serves the role of sourcing the geometry, whereas the Brown-York stress-energy tensor describes the properties of a hypersurface within the geometry. More specifically, the Brown-York stress tensor provides a useful prescription for computing the energy and momentum density of a cutoff surface within a gravitational system. In the present case, the Brown-York stress tensor relates the energy and momentum on a cutoff surface $\Sigma_c$ to the energy and moment of a fluid. We also note that for the simple fluid \eqref{eq:perfect fluid}, the spacetime stress-energy tensor $\mathcal{T}_{\mu\nu}$ is clearly zero, since the spacetime is just empty Minkowski space.  By contrast, the fluid does have a nonzero Brown-York tensor as we have shown.

\subsection{The Hydrodynamic Limit and Near-Horizon Expansion}
\label{sub:hydro limit}
In order to obtain non-constant fluids, the next step is to perturb the seed metric \eqref{eq:seed metric} by allowing the fluid degrees of freedom $u_a, p$ to depend on $(\tau, x^i)$, following the procedure in~\cite{Compere:2011dx,Compere:2012mt, Pinzani-Fokeeva:2014cka}.  Then, near-equilibrium solutions $g^{(n)}_{\mu\nu}$ ($n=0,1,2\dots$) can be constructed order-by-order in the relativistic gradient expansion. However, in this work we are interested in the non-relativistic hydrodynamic limit since it agrees with the near-horizon expansion~\cite{Bredberg:2011jq,Keeler:2020rcv}, and we will be working in the near-horizon regime for the Rindler shockwaves below.

 Accordingly, in terms of the nonrelativistic fluid velocity $v_i$, and the nonrelativistic pressure $P$, we have
\begin{align}\label{eq:velocity}
    u_a=\frac{1}{\sqrt{\alpha r_c-v^2}}\left(-\alpha r_c, v_i\right), \qquad p = (\alpha r_c -2 P)^{-1/2},
\end{align}
where the relativistic velocity is normalized so $u_a u^a=-1$. 
The equivalent of the relativistic gradient expansion becomes explicitly the hydrodynamic scaling~
\begin{equation} \label{eq:hydro scaling}
	v_i \sim \order{\e}, \quad P \sim \order{\e^2}, \quad \pd_i \sim \order{\e}, \quad \pd_{\tau} \sim \order{\e^2}.
\end{equation}
Using this non-relativistic scaling, and further solving the non-constraint Einstein equations order-by-order, we construct the non-relativistic fluid metric given in Appendix~\eqref{eq:hydro metric}.  Although the full fluid-dual geometry is quite complicated, with its metric given by~\eqref{eq:hydro metric} up to $\order{\varepsilon^3}$, we are mainly interested in the lowest order expansion of~\eqref{eq:hydro metric}:%
\footnote{We ignore $\order{\varepsilon^2}$ and higher order terms presently because we anticipate the relation between the fluid-dual metric in just the leading order of hydrodynamic expansion~\eqref{eq:linear fluid metric} and the shockwave geometry that solves the linearized Einstein equation. This point will become clear in section~\ref{sec:shockwaves}.}%
\begin{align} \label{eq:linear fluid metric}
	\begin{split}
		d s^2 = &-(\a r) d \tau^2 + 2 d \tau d r + \d_{i j} d x^i d x^j \\
					&- 2 \qty(1 - \frac{r}{r_c}) v_i d x^i d \tau - 2 \frac{v_i}{\a r_c} d x^i d r + \order{\varepsilon^2}.
	\end{split}
\end{align}

As noted above, the fluid behavior is described via the Brown-York stress tensor~\eqref{eq:Tab} and its conservation equations. Explicitly in terms of the non-relativistic fluid metric, we obtain~\cite{Bredberg:2011jq}%
\footnote{We used the symmetrization $X_{(i} Y_{j)}=\frac{1}{2}\left(X_iY_j+X_j Y_i\right).$}
\begin{equation} \label{eq:Tab explicit}
	\kappa^2\, T_{a b} d x^a d x^b = \frac{\a v^2}{\sqrt{\a r_c}} d \tau^2  - \frac{2\a v_i}{\sqrt{\a r_c}} d x^i d \tau +\frac{\a}{\sqrt{\a r_c}} d x^i d x_i + \frac{P \delta_{i j} + v_i v_j }{r_c \sqrt{\a r_c}} d x^i d x^j - \frac{2 \pd_{(i} v_{j)} }{\sqrt{\a r_c}} d x^i d x^j + \order{\e^3}.
\end{equation}
where details of the derivation are included in Appendix~\ref{app:BY Derivation}.

Since the constraint Einstein equations on $\Sigma_c$ match the conservation equations for the Brown-York tensor, we find\footnote{This is the Gauss-Codazzi equation including the integration constant in $r$ \cite{Pinzani-Fokeeva:2014cka}. We set the integration constant to zero by imposing the Brown-York stress tensor to be conserved when we consider the source-free situation.} 
\begin{equation} \label{eq:conserv BY tensor}
	\nabla^b T_{a b} |_{\Sigma_c} = -2 n^{\mu} \mathcal{T}_{a \mu}\,.
\end{equation}
As before, $n^\mu$ is the unit normal vector~\eqref{eq:unit normal} to the cutoff surface $\Sigma_c$.  The spacetime stress-energy tensor $\mathcal{T}_{a \mu}$ provides the nonzero source for Einstein's equations, and we again wish to differentiate it from the Brown-York tensor we introduced in Sec.~\ref{sec:fluids setup}. 

In section \ref{sec:shockwaves}, we are interested in the spacetime stress-energy tensor $\mathcal{T}_{\mu\nu}$  due to an ultra-relativistic source, also known as the gravitational shockwave. Before turning on the external sources, we now review the fluid behavior when no source is present, that is, when  $\mathcal{T}_{\mu\nu}=0$. At $\order{\e^0}$, $T_{a b}$ is constant, and the Einstein equations are satisfied trivially. At $\order{\e^2}$, we obtain the incompressibility condition~\cite{Bredberg:2011jq} 
\begin{equation} \label{eq:conserv BY tensor incompress}
	\nabla^a T_{\tau a}=0 \quad \iff \quad G_{\tau \tau} = \frac{\alpha}{2}\pd^i v_i = 0\,.
\end{equation}
In writing out~\eqref{eq:conserv BY tensor incompress}, we have identified the conservation of the Brown-York tensor at $\order{\varepsilon^2}$ with the Einstein tensor $G_{\tau \tau}$ at the same order~\cite{Pinzani-Fokeeva:2014cka}. We note that other components of the Einstein tensor vanish identically at the current order.

In a similar fashion, we compute the conservation equation of the Brown-York tensor at $\order{\e^3}$, and we obtain the incompressible Navier-Stokes equation~\cite{Bredberg:2011jq}
\begin{equation}\label{eq:navier stokes}
	\nabla^a T_{i a}=0 \quad \iff  \quad G_{\tau i} = -\frac{1}{2 r_c} \qty(\pd_{\tau} v_i - r_c \pd^2 v_i + \pd_i P + v_j \pd^j v_i) = 0,
\end{equation}
which is equivalent to the vacuum constraint Einstein equation $G_{\tau i}=0$ evaluated at the same order~\cite{Pinzani-Fokeeva:2014cka}.

Later, when we add sources to the Einstein equations, these constraint equations will instead become \eqref{eq:conserv BY tensor}, adding sources to the fluid equations.
In section \ref{sec:shockwaves}, we will see how a nonzero $\mathcal{T}_{\mu\nu}$ coming from gravitational shockwaves influences the fluid on $\Sigma_c$.

%%%%%%%%%%
\paragraph{Potential Fluids} 

Before considering shockwave spacetimes, we review potential fluids.  The class of fluids known as potential fluids satisfies 
\begin{align}
    v_i=\partial_i \phi\,.
\end{align} 
These fluids are also known as irrotational, since they have zero vorticity.  In our 4-dimensional gravity (dual to a 2+1 dimensional fluid), the vorticity is $\partial_x v_y-\partial_y v_x=0$, since gradients are curl-free.

For a source-free potential fluid, the incompressibility condition \eqref{eq:conserv BY tensor incompress} yields~\cite{Keeler:2020rcv}
\begin{align}\label{eq:incomp potential wo source}
    \nabla^a T_{0 a}=0 \quad\imp\quad \partial^2 \phi = 0\,,
\end{align}
In this case, the incompressible Navier-Stokes equation \eqref{eq:navier stokes} implies the expression of the pressure in terms of $\phi$
\begin{align}\label{eq:NS potential wo source}
    \nabla^a T_{i a}=0 \quad\imp \quad \partial_i P=-\partial_i \partial_\tau \phi-\partial^j \phi \partial_i \partial_j \phi\,.
\end{align}

%%%%%%%%%%%%%%%%%%%%%%
\paragraph{Rindler to Light Cone Coordinates}

Lastly, we review the coordinate transformation between the background to the fluid-gravity dual metrics, and light cone coordinates.  We review this relationship because both the fluid-dual metrics, and the shockwave metrics we study in the next section, are perturbations away from Minkowski space.  However in the fluid case we work in ingoing Eddington-Finkelstein coordinates \eqref{eq:ingoing rindler}, while in the shockwave case the background is written in light cone coordinates.

For the shockwave, the background is Minkowski spacetime in light-cone coordinates $u=T+Z,\, v=T-Z$ where $(T,Z)$ are Minkowski global coordinates, with metric
\begin{align}\label{eq:Minkowski}
    ds^2 = -du dv + \delta_{ij}dx^i dx^j.
\end{align}
Here again $i,j$ indicate the transverse directions. The coordinate transformation
\begin{equation} \label{eq:u v diffeo}
	u = \sqrt{ \frac{r_c}{\a} } e^{\a \tau/ 2} \qand   v = - \frac{4 r}{ \sqrt{\a r_c}} e^{- \a \tau/ 2}\,
\end{equation}
yields the line element of Rindler spacetime \eqref{eq:ingoing rindler}. Thus, Eq.~\eqref{eq:u v diffeo} is the zeroth order coordinate transformation between the fluid metric~\eqref{eq:linear fluid metric} and shockwave metric~\eqref{eq:AS metric}. In the next section, we construct the relation at the next order, between the perturbed geometries.

%%%%%%%%%%%%%%%%%%%%%%

\section{Near Horizon Fluids from Gravitational Shockwaves}
\label{sec:shockwaves}
In this section, we establish the relation between the near-horizon fluids constructed in section~\ref{sub:hydro limit} and the gravitational shockwaves in \cite{ASmetric}. We first briefly review the shockwave geometry and the external source from the shockwave. Then, we consider how the fluid equations deforms in the presence of an external source. With those two setups constructed, we find the connection between the fluids and shockwaves by using their Petrov classifications.

\subsection{Gravitational shockwave geometry}

The gravitational shockwave geometry was originally proposed by Aichelburg and Sexl~\cite{ASmetric} to describe gravitational radiation from a massless point particle. The Aichelburg-Sexl metric is given by~\cite{DRAY1985173, Verlinde_2022} with Minkowski spacetime in the light-cone coordinates \eqref{eq:Minkowski}.

If we consider a fast particle falling towards the future horizon, the $uu-$component of the metric is perturbed, which produces the shockwave geometry 
\begin{equation} \label{eq:AS metric}
	d s^2 = - du d v + H_{uu} du^2 + \d_{i j}d x^i d x^j.
\end{equation}
Here $H_{uu}$ is given by 
\begin{equation} \label{eq:Huu shock}
	H_{uu}(u, x^i) = p_u \kappa^2 \delta(u - u_0) f(\bm{x}; \bm{x}'),
\end{equation}
where $p_u$ is the constant momentum of the shock, $u_0$ denotes the location of the shockwave, and $f(\bm{x}; \bm{x}')$ is the Green's function of the transverse Laplacian operator:
\begin{equation} \label{eq:Green function laplacian}
	- \pd^2 f(\bm{x}; \bm{x}') = \delta^2(\bm{x} - \bm{x}').
\end{equation}
In fact, Eq.~\eqref{eq:Green function laplacian} is equivalent to the Einstein equation, which is exactly solvable in this case.  Only the $uu$ component of the Einstein equation $G_{\mu\nu} = \kappa^2 \mathcal{T}_{\mu\nu}$ is nontrivial. Explicitly, this component becomes
\begin{equation} \label{eq:shockwave einstein eq}
    - \frac{1}{2} \pd^2 H_{uu} = \kappa^2\mathcal{T}_{uu},
\end{equation}
where $\mathcal{T}_{uu}$ is the only non-vanishing component of the shockwave stress-energy tensor.  For $H_{uu}$ as in \eqref{eq:Huu shock}, we have
\begin{equation} \label{eq:Tuu shock}
	\mathcal{T}_{uu} = \frac{1}{2}p_u \delta(u - u_0) \delta^{2}(\bm{x}- \bm{x}').
\end{equation}
The shockwave geometry can be described in a different gauge, which involves an off-diagonal component with transverse and longitudinal directions. The metric in this gauge is
\begin{align}\label{eq:mixed gauge}
    d s^2=-d u d v+ \pd_i X^v d u d x^i+\delta_{ij}dx^i dx^j
\end{align}
where we introduced the shockwave degree of freedom $X^v$
\begin{align}\label{eq:classical Xv}
    X^v(u, v, x^i)=v+\int_{0}^u d u^{\prime} H_{u u}\left(u^{\prime}, x^i\right).
\end{align}
Now, we have the shockwave geometry sourced by the external matter stress-energy tensor $\mathcal{T}_{uu}$. In the next section, we will construct a fluid-dual geometry with the same external source in order to make a connection between shockwaves and fluids.

\subsection{Near Horizon Fluids with External Sources}
\label{sub:diffeo shock and fluid}

Just as the stress-energy tensor $\mathcal{T}_{uu}$ in \eqref{eq:Tuu shock} sources the shockwave geometry with nonzero $H_{uu}$, in fluid-dual geometries an external stress-energy tensor can be thought of as a sourcing the fluid dynamics. We anticipate that in the leading order of hydrodynamic expansion the fluid-dual metric given by Eq.~\eqref{eq:linear fluid metric} can be mapped to the shockwave geometry given by Eq.~\eqref{eq:AS metric}. In anticipation of this match, we first add a non-zero external source $\mathcal{T}_{uu}$ to the fluid equations~\eqref{eq:conserv BY tensor}, generalizing the fluid/gravity setup in \cite{Bredberg:2011jq} beyond vacuum solutions. 

Since the shockwave metric and fluid metric use different coordinate systems, we need to find the components of $\mathcal{T}_{\mu\nu}$ from \eqref{eq:Tuu shock} in $(\tau,r,x^i)$ coordinates. 
Since the only non-vanishing component of $\mathcal{T}_{\mu\nu}$ in $(u,v, x^i)$ coordinates is $\mathcal{T}_{u u}$, under the background coordinate transformation~\eqref{eq:u v diffeo}
we find that the only non-vanishing component in $(\tau,r, x^i)$ coordinates is $\mathcal{T}_{\tau \tau}$, given by
\begin{equation} \label{eq:T tau tau}
	\mathcal{T}_{\tau \tau} =\frac{\a r_c}{4} e^{\a \tau}  \mathcal{T}_{uu}.
\end{equation}

In Sec.~\ref{sec:fluids setup}, we showed that the conservation Eq.~\eqref{eq:conserv BY tensor} of the Brown-York stress tensor at order $\order{\e^2}$  is equivalent to the incompressibility condition of the dual fluid~\eqref{eq:conserv BY tensor incompress}. However, in the presence of a gravitational shockwave, the incompressibility condition instead becomes
\begin{equation} \label{eq:incomp shock}
    \nabla^a T_{\tau a} |_{\Sigma_c}= -2  n^{\mu} \mathcal{T}_{\tau \mu} \quad\imp\quad \pd^i v_i = \frac{1}{2}\kappa^2 r_c e^{\a \tau} \mathcal{T}_{uu},
\end{equation}
where we have additionally made use of the unit normal as explicitly given in \eqref{eq:unit normal}.

At $\mathcal{O}(\ep^3)$, the conservation Eq.~\eqref{eq:conserv BY tensor} of the Brown-York stress tensor is given by
\begin{align}\label{eq:NS shock}
    \nabla^a T_{i a} |_{\Sigma_c}=-2 n^{\mu} \mathcal{T}_{i \mu} \imp \partial_{\tau} v_i-r_c \partial^2 v_i+v_j \partial^j v_i+\partial_i P+v_i \partial^j v_j-r_c \partial_i \partial^j v_j=0.
\end{align}
There is no contribution from the external source since the only non-vanishing component of $\mathcal{T}_{\mu\nu}$ in $(\tau,r,x^i)$ coordinates is $\mathcal{T}_{\tau\tau}$. However, the equation is not equivalent to the incompressible Navier-Stokes equation \eqref{eq:navier stokes}; instead, it has acquired one further term: $r_c \pd_i \pd^j v_j$, since incompressibility is broken.

\paragraph{Potential Fluids} If we assume a potential fluid of the form $v_i=\partial_i \phi$, the conservation equation of the Brown-York stress tensor in the leading order \eqref{eq:incomp shock} becomes
\begin{align}\label{eq:incomp potential}
\nabla^a T_{\tau a} |_{\Sigma_c}= -2 n^{\mu} \mathcal{T}_{\tau \mu}  \quad\imp\quad \partial^2 \phi=\frac{1}{2}\kappa^2 r_c e^{\a \tau} \mathcal{T}_{uu}.\,
\end{align}
Upon examining Eq.~\eqref{eq:incomp potential} closely with the Einstein equation of the shockwave geometry~\eqref{eq:shockwave einstein eq}, we are tempted to believe that the two geometries are equivalent, at least in the leading order of hydrodynamic expansion. Specifically, we compare Eqs.~\eqref{eq:incomp potential} and~\eqref{eq:shockwave einstein eq} term by term
\begin{equation} \label{eq:EE compare}
	\pd^2 \phi = \frac{1}{2} \kappa^2 r_c e^{\a \tau} \mathcal{T}_{uu} \quad \stackrel{?}{\iff}  \quad - \frac{1}{2} \pd^2 H_{uu} = \kappa^2 \mathcal{T}_{uu}
\end{equation}
and deduce a relation between $H_{uu}$ and $\phi$ to be
\begin{equation} \label{eq:phi and Huu}
	\phi = - \frac{r_c}{4} e^{\a \tau} H_{uu}.
\end{equation}
While this may suggest a mapping between a potential fluid and the shockwave metric, it does not guarantee the two metrics are diffeomorphic. The most reassuring way to see whether two spacetimes are equivalent is to construct an explicit diffeomorphism that maps one to the other. However, in our case, finding such a diffeomorphism is quite difficult (considering the form of the fluid metric~\eqref{eq:linear fluid metric}), if not impossible. Thankfully, there is a more straightforward method to connect the shockwave geometry with the potential fluid, because both geometries have Weyl tensors that exhibit explicit algebraic speciality~\cite{stephaniExactSolutionsEinstein2003,Keeler:2020rcv}. In Sec.~\ref{sub:petrov}, we utilize the Petrov classification to find a correspondence between the shockwave and potential fluid geometries, thereby substantiating our claim in Eq.~\eqref{eq:phi and Huu}.

Last but not least, at $\mathcal{O}(\ep^3)$, the conservation equation~\eqref{eq:NS shock} becomes
\begin{align}\label{eq:NS potential}
    \nabla^a T_{i a} |_{\Sigma_c}=-2 n^{\mu} \mathcal{T}_{i \mu} \quad\imp \quad\partial_{\tau} \partial_i \phi-2 r_c \partial_i \partial^2 \phi + \pd_i P +\partial_j \phi \,\partial^j \partial_i \phi + \partial_i \phi \,\partial^2\phi=0.
\end{align}
Eqs.~\eqref{eq:incomp shock} and \eqref{eq:NS shock} determine fluid dynamics on the cutoff surface when we have added the shockwave stress-energy tensor $\mathcal{T}_{uu}$ as an external source. Since the shockwave geometry~\eqref{eq:mixed gauge} and the fluid-dual geometry are both perturbations away from Minkowski spacetime, when we set them to have the same external source, we expect that they are two equivalent geometries in different gauges. In the next section, we establish the explicit connection between the shockwave metric and the fluid-dual metric.

\subsection{Petrov Classification Connecting Fluids with Shockwaves}
\label{sub:petrov}
In this section, we derive the relation between the near-horizon fluid metric~\eqref{eq:linear fluid metric} and the shockwave metric~\eqref{eq:AS metric}. Instead of finding the complicated diffeomorphism between them, we will adopt the virtue of the Newman-Penrose formalism to link the shockwave degrees of freedom into the fluid story. We will follow the conventions of \cite{stephaniExactSolutionsEinstein2003} throughout.

The Newman-Penrose formalism will allow us to calculate the Petrov classification of our spacetimes. The Petrov classification categorizes spacetimes according to the multiplicities of the principal null directions of the Weyl tensor $C_{abcd}$. Principal null directions $k^\mu$ are null vectors satisfying
\begin{align}\label{eq:principal null condition}
    k_{[\mu} C_{\nu] \rho \sigma[\gamma} k_{\delta]} k^\rho k^\sigma=0\,.
\end{align}

All four-dimensional spacetimes have four principal null vectors. Spacetimes with four distinct principal null vectors for the Weyl tensor are not algebraically special; they are said to be of Petrov type I. If two principal null directions coincide, then the spacetime is Petrov type II. If the four principal nulls coincide in pairs, then the spacetime is type D. If all four principal null directions coincide, then the spacetime is type N.

Moreover, if the spacetime is type N, the geometry is determined by only one degree of freedom as we will see explicitly in this section. As we will review below, both the shockwave metric and the metric dual to a potential fluid are type N, so we can match them by matching the single degree of freedom. 

We begin with the shockwave metric, defining the complex coordinates
\begin{align}\label{eq:complex coordi}
    z=x+iy,\quad \bar{z}=x-iy,
\end{align}
where we are now explicitly restricting to four-dimensional spacetime. Then, the shockwave metric \eqref{eq:AS metric} becomes
\begin{align}\label{eq:complex shock}
    d s^2= -d u d v+H_{uu}(u,z,\bar{z}) d u^2+dz d\bar{z}\,.
\end{align}
In the Newman-Penrose formalism, the metric is rewritten in terms of a null tetrad, $(k_\mu,\, l_\mu,\,m_\mu,\,\bar{m}_\mu)$.  Explicity, $g_{\mu\nu} = -l_{(\mu}n_{\nu)}+m_{(\mu}\bar{m}_{\nu)}$.
For the shockwave geometry, we will use the null tetrad
\begin{align}\label{eq:shock null tetrads}
m^\mu\partial_\mu=\sqrt{2}\partial_z, \quad \bar{m}^\mu\partial_\mu=\sqrt{2}\partial_{\bar{z}}, \quad l^\mu\partial_\mu=\sqrt{2}\partial_u-\sqrt{2} H_{uu} \partial_v, \quad k^\mu\partial_\mu=\sqrt{2} \partial_v\,.
\end{align}
Here $k^\mu$ is a principal null direction since it satisfies \eqref{eq:principal null condition}. Indeed, since
$k^\mu$ obeys the condition
\begin{align}\label{eq:type N condition}
    C_{\mu\nu\rho\sigma} k^\sigma=0,
\end{align}
it is a 4-fold repeated principal null vector, so the shockwave geometry is type N.  Importantly, when the null tetrad is chosen so that the first vector in the null tetrad is itself a 4-fold repeated principal null, the Weyl scalars
 $\Psi_i, i=0, \ldots, 4$ take on a special form, namely $\Psi_0=\Psi_1=\Psi_2=\Psi_3=0, \Psi_4 \neq 0$.
  
 In the tetrad formalism, these Weyl scalars contain all the information in the Weyl tensor.  In general,
\begin{align}\label{eq:Weyl scalars}
\begin{array}{ll}
\Psi_0 \equiv C_{\mu\nu\rho\sigma} k^\mu m^\nu k^\rho m^\sigma,\quad & \Psi_1 \equiv C_{\mu\nu\rho\sigma} k^\mu l^\nu k^\rho m^\sigma, \quad \Psi_2 \equiv C_{\mu\nu\rho\sigma} k^\mu m^\nu \bar{m}^\rho l^\sigma\,\\
\Psi_3 \equiv C_{\mu\nu\rho\sigma} k^\mu l^\nu \bar{m}^\rho l^\sigma, & \Psi_4 \equiv C_{\mu\nu\rho\sigma} \bar{m}^\mu l^\nu \bar{m}^\rho l^\sigma\,.
\end{array}
\end{align}
These scalars do depend on the choice of null tetrad, but if there is any null tetrad in which only $\Psi_4$ is nonzero, then the spacetime must be type N.

For fluid dual metrics, the non-relativistic fluid geometry \eqref{eq:hydro metric} is algebraically special, specifically of type II, as first shown in \cite{Bredberg:2011jq}. More recently, \cite{Keeler:2020rcv} showed that two special types of fluid have higher algebraic speciality. For a constant vorticity fluid, the fluid-dual metric is type D, while for a potential fluid it is type N. 

Since both the fluid-dual metric for a potential fluid, and the shockwave geometry \eqref{eq:AS metric}, are type N, their Weyl tensors are described by the single degree of freedom $\Psi_4$ (in a tetrad choice where all other components are zero). Indeed, a potential fluid is described by only the scalar $\phi$, while a shockwave is described by the single function $H_{uu}$, so we should not be too surprised that they are both type N.

In order to relate the shockwave geometry \eqref{eq:AS metric} and the fluid metric \eqref{eq:linear fluid metric}, we will choose a tetrad for each where only $\Psi_4$ is nonzero. If two Petrov type N geometries actually represent the same spacetime, then if tetrads are chosen so that all other $\Psi_i$ vanish, the $\Psi_4$'s for the two geometries must match up to an overall scaling.

For the shockwave metric \eqref{eq:AS metric}, the non-vanishing Weyl scalar is given by
\begin{align}\label{eq:Psi4 shock}
    \Psi_4= 2\partial_{\bar{z}} \partial_{\bar{z}}H_{uu}(u,z, \bar{z})\,.
\end{align}
For the fluid metric \eqref{eq:linear fluid metric} dual to a potential fluid, the tetrad can be chosen so the non-vanishing Weyl scalar is given by \cite{Keeler:2020rcv}
\begin{align}\label{eq:Psi4 fluid}
    \Psi_4=\frac{2}{r} \partial_{\bar{z}}\partial_{\bar{z}} \phi (\tau, z,\bar{z})\,.
\end{align}
Indeed, we find that we can relate the potential fluid $\phi$ to the shockwave metric component $H_{uu}$ by equating these two $\Psi_4$'s, up to an overall scale $\beta$:
\begin{align}\label{eq:phi and H}
    \phi=\beta r H_{uu}\,.
\end{align}

Note that we have one more constraint when we relate the fluid degrees of freedom with the shockwave one: the leading-order conservation equation of the Brown-York stress tensor \eqref{eq:incomp shock} should be consistent with the Einstein equation \eqref{eq:shockwave einstein eq}. This additional constraint enables us to determine the overall scaling ambiguity $\beta$. Inserting~\eqref{eq:phi and H} into the leading-order fluid equation \eqref{eq:incomp shock}, we obtain
\begin{equation} \label{eq:fluid shock eom}
    \beta r \pd^2 H_{uu} = \frac{1}{2}\kappa^2 r_c e^{\a \tau} \mathcal{T}_{uu}.
\end{equation}
By judiciously choosing the value of $\beta$ to be 

\begin{equation} \label{eq:beta}
	\beta = - \frac{r_c e^{\a \tau}}{4 r} = \frac{u}{v},
\end{equation}
we discover that $\order{\varepsilon^2}$ conservation equation in the fluid-dual metric matches exactly with the shockwave equation of motion~\eqref{eq:shockwave einstein eq}. 

In terms of the shockwave degree of freedom $X^v$ \eqref{eq:classical Xv}, the conservation equations of the Brown-York stress tensor become
\begin{align}
& \order{\varepsilon^2}:\qquad -\frac{1}{2}\partial^2 \partial_u X^v=\kappa^2 \mathcal{T}_{uu}\label{eq:second-order eq} \\
& \order{\varepsilon^3}:\qquad \alpha u^2 \pd_i\partial_u^2 X^v+2\alpha u \pd_i \partial_u X^v-4 r_c u  \pd_i \partial_u \partial^2 X^v  \approx \frac{8}{\alpha u} \pd_iP \label{eq:third-order eq}
\end{align}
 In the second line, we have used $\partial_\tau H_{uu}=\frac{\alpha}{2} u \partial_u H_{uu}$, since $\partial_v H_{uu}=0$.  Additionally, we have ignored terms with $X_v$ appearing twice.  Regardless, the $\mathcal{O}(\e^3)$ equation here has no contribution from the stress-energy tensor sourced by the shockwave, as we already anticipated in \eqref{eq:NS potential}. Note that the $\mathcal{O}(\e^2)$ equation contributes to the third term in the $\mathcal{O}(\e^3)$ equation, leading to
\begin{align}\label{eq:third-order in time}
    \order{\varepsilon^3}:\qquad 
    2 e^{-\alpha\tau/2} P \approx \sqrt{\frac{r_c}{\alpha}}\partial_\tau^2 X^v + \frac{\sqrt{\alpha r_c}}{2}\partial_\tau X^v + 2(\kappa r_c)^2 e^{\alpha\tau/2}\mathcal{T}_{uu},
\end{align}
where we used $(\tau,r,x^i)$ coordinates, and the external source $\mathcal{T}_{uu}$ for the shockwave \eqref{eq:complex shock} is given by~\eqref{eq:Tuu shock}. Note that we have omitted nonlinear coupling terms in~\eqref{eq:third-order eq} and~\eqref{eq:third-order in time}. Because time derivatives do not act on the nonlinear coupling terms, coordinate transformation from $(u,v, x^i)$ to $(\tau,r, x^i)$ coordinates leave them invariant.

So far, all of the calculations are performed at the classical level. However, by instead promoting our source to a quantum operator, we will find a structure quite similar to \cite{Verlinde_2022,Zhang:2023mkf}.  Specifically, we introduce quantum light-ray operators associated with the shockwave degrees of freedom and impose the t'Hooft commutation relation. Our goal is to study the two-point correlation functions on the cutoff hypersurface $\Sigma_c$, which is timelike and boost invariant. In the next section, we will thus generalize the spacetime fluctuations on the null hypersurface \cite{Verlinde_2022} to a timelike hypersurface, by utilizing the cutoff fluid approach that we constructed in the previous sections.

\section{Spacetime Fluctuations from Quantum Sources}
\label{sec:fluctuation dissipation}

In this section, we explore how spacetime fluctuations from quantum sources influence the behavior of fluids living on $\Sigma_c$. In particular, we will utilize the 't Hooft commutation relations~\cite{tHooft:1996rdg} to modify the classical incompressibility equation to a stochastic differential equation. Then we follow the procedure in~\cite{Zhang:2023mkf} to solve the differential equation and to compute the variance in photon traversal time.

Our critical assumption is that due to vacuum energy fluctuations, $\mathcal{T}_{uu}$ is no longer the stress-energy tensor of a classical shockwave. Rather, $\mathcal{T}_{uu}$ encapsulates features of quantum fluctuations. We first utilize a commutation relation between $H_{vv}$ and $\mathcal{T}_{uu}$, in close analogy to those proposed by 't Hooft~\cite{Verlinde_2022,Zhang:2023mkf}:
\begin{equation} \label{eq:comm T H unequal time}
	\comm{\mathcal{T}_{uu}(x)}{H_{vv}(x')} = i \d^4(x-x'),
\end{equation}
where $x$ denotes coordinates in full spacetime dimensions. In writing Eq.~\eqref{eq:comm T H unequal time} we have assumed that $\mathcal{T}_{uu}$ and $H_{vv}$ are quantum operators. 

\begin{figure}
	\centering
	\hspace*{-0.5in}
	\includegraphics[width=0.5\textwidth]{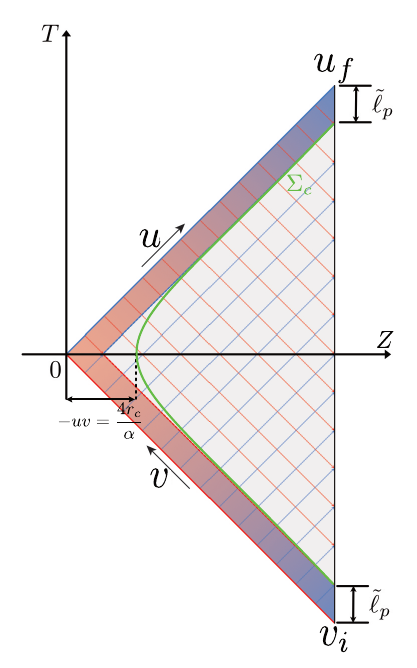} 
	\vspace*{-3mm}
	\caption{Minkowski Rindler space that represents a slice of a Minkowski diamond of size $L$. Blue and red dashed lines represent a series of nested causal diamonds along the past (future) light front. The shaded blue/red region is the ``smearing'' of the diamond due to quantum fluctuations modeled by gravitational shockwaves. This procedure is discussed in detail in Ref.~\cite{Zhang:2023mkf}. In this paper, we aim to describe the fuzzing of the light front in a Lorentz invariant manner. This leads to a hyperbolic cutoff surface (green curve), denoted as $\Sigma_c$. The proper distance between $\Sigma_c$ and the bifurcate Rindler horizon is given by the relation $- u v = 4 r_c / \a$.}  \label{fig:cutoff}
\end{figure}

Here we have introduced an additional degree of freedom $H_{vv}$, which classically lives only on the past horizon. Although it is true that along either the future or past light front, only $H_{uu}$ or $H_{vv}$ is non-vanishing classically, in the present case $\mathcal{T}_{uu}$ and $H_{vv}$ act as quantum mechanical conjugate operators, as discussed in Refs.~\cite{Verlinde_2022,He:2023qha}. Their commutation relation~\eqref{eq:comm T H unequal time} is formally evaluated at the bifurcate horizon. Nevertheless, we can heuristically argue~\cite{Banks:2021jwj,Zurek:2022xzl,Zhang:2023mkf} that a single causal diamond is foliated by a series of nested causal diamonds (see Fig.~\ref{fig:cutoff}), each separated by a microscopic length scale $\tilde{\ell}_p$, known as the decoherence length. Beyond this length scale, the subsequent diamonds become uncorrelated~\cite{Banks:2021jwj,Zurek:2022xzl,Zhang:2023mkf}. This argument allows us to identify a series of bifurcate horizons along the future and past light front, enabling us to introduce $H_{vv}$ as an ultra local quantum degree-of-freedom. Crucially, the notion of a nested causal diamond gives rise to an accumulated quantum uncertainty in the photon traversal time~\cite{Banks:2021jwj,Zurek:2022xzl,Zhang:2023mkf}. Next, we proceed to compute this uncertainty in detailed steps. 

 Applying the commutator~\eqref{eq:comm T H unequal time}, we immediately discover that the classical shockwave equation of motion~\eqref{eq:shockwave einstein eq}, or equivalently, the incompressibility equation for fluids~\eqref{eq:incomp potential} is now a differential equation involving quantum commutators 
\begin{equation} \label{eq:comm Huu Hvv laplacian}
	-\pd^2 \comm{H_{uu}(u, \bm{x})}{H_{vv}(v, \bm{x}')} = i \ell_p^2 \d(u - u_0) \d(v -v_0) \d^2(\bm{x} - \bm{x}'),
\end{equation}
where we have expressed $\d^4(x - x')$ in the $(u,v, x^i)$ coordinates and expressed $\kappa^2 = \ell_p^2$ explicitly. Note that the operator $\pd^2$ only acts on the un-primed quantity $H_{uu}$. Eq.~\eqref{eq:comm Huu Hvv laplacian} is the Minkowski limit of the AdS spacetime studied in Ref.~\cite{Zhang:2023mkf}. By going through an analogous procedure, we obtain the following commutation relations 
\begin{equation} \label{eq:comm Huu Hvv}
	\comm{H_{uu}(u, \bm{x})}{H_{vv}(v, \bm{x}')} = i\ell_p^2 \d(u - u_0) \d(v - v_0) f(\bm{x};\bm{x}'), 
\end{equation}
where $f(\bm{x}, \bm{x}')$ is given by Eq.~\eqref{eq:Green function laplacian}.

 Because $\mathcal{T}_{uu}$ is stochastic, it naturally implies that $\expval{H_{uu}}$ and $\expval{H_{vv}}$ vanish. However, the variance $\expval{H_{uu}^{2}}$ (similarly for $u \to v$) in general does not vanish. To compute the variance, we invoke the Robertson uncertainty principle~\cite{Verlinde_2022,Zhang:2023mkf}
\begin{equation} \label{eq:Robertson uncertainty}
	\expval{H_{uu}^2} \expval{H_{vv}^{2}} \ge \abs{\frac{1}{2 i} \expval{\comm{H_{uu}}{H_{vv}}}}^2 = \qty(\frac{\ell_p^2}{2})^2 \qty[\d(u -u_0) \d(v - v_0) f(\bm{x}; \bm{x}')]^2.
\end{equation}
Assuming that both are minimum uncertainty states, then the quantities $\expval{H_{uu}^{2}}$ and $\expval{H_{vv}^{2}}$ are equal:
\begin{equation} \label{eq:var Huu Hvv min uncert}
\expval{H_{uu}^2} = \expval{H_{vv}^{2}} = \frac{\ell_p^2}{2} \d(u - u_0) \d(v - v_0) f (\bm{x}; \bm{x}').
\end{equation}
In order to further evaluate $\expval{H_{uu}^{2}}$ and $\expval{H_{vv}^{2}}$, we follow the heuristic argument of nested causal diamonds and statistical independence once again. This line of reasoning follows precisely the logic presented in details in Refs.~\cite{Banks:2021jwj,Zhang:2023mkf}, and we will not reproduce it in this paper.

However, one crucial distinction sets our current paper apart from~\cite{Zhang:2023mkf}. With the help of the cutoff fluid approach, we implement a ``smeared-out'' horizon (Fig.~\ref{fig:cutoff}), using $\Sigma_c$ so we can regularize the appropriate delta function in a manifestly Lorentz invariant manner. The first step in our regularization procedure is to transform the lightcone coordinates $(u,v, x^i)$ into the ingoing coordinates $(\tau,r, x^i)$ naturally adapted to describe $\Sigma_c$. The two dimensional delta function in the $(u,v)$ coordinates are written explicitly in the $(\tau, r)$ coordinates as:
\begin{align}
	\d(\tau(u) - \tau_0) &= \frac{\d(u - u_0)}{\abs{\tau'(u)}} = \frac{\a u}{2} \d(u - u_0),\label{eq:delta tau tau0} \\
	\delta (r - r_0) &= \delta\qty(-\frac{\a u v}{4} - r_0) = \delta(\a u v / 4 + r_0) = \frac{4}{\a u}\d(v - v_0), \label{eq:delta r r0}
\end{align}
where $\tau'(u) = \pdv*{\tau}{u}$. The function $\d(\tau - \tau_0)$ keeps track of the causal development of an observer traversing on $\Sigma_c$, while $\delta( r - r_0)$ sets the width of the cutoff surface $\Sigma_c$. Therefore, we will replace $\delta(r - r_0)$ with a suitably regularized kernel. After first rescaling the argument of $\delta (r - r_0)$ we obtain
\begin{equation} \label{eq:rescale delta}
	\d(r - r_0) = \d \qty(\frac{\a u v}{4} + r_0) = \frac{4}{\a} \d\qty(u v + \frac{4 r_0}{\a}).
\end{equation}
Next, we regularize the delta function with a Poisson kernel\footnote{The delta function can be represented by various kernels, but choosing the Poisson kernel allows us to keep consistency with the previous work~\cite{Zhang:2023mkf}.} 
\begin{equation} \label{eq:poisson kernel}
	\d \qty(u v + \frac{4 r_0}{\a}) = \lim_{\epsilon \to 0} \frac{2}{\pi} \frac{\epsilon}{\epsilon^2 + (uv + 4 r_0/\a)^2} \approx \frac{2}{\pi} \frac{1}{\epsilon}.
\end{equation}
In obtaining the last quantity, we have evaluated Eq.~\eqref{eq:poisson kernel} on $\Sigma_c$, where $r_0 = r_c$ and $uv$ satisfies the relation
\begin{equation} \label{eq:uv cutoff}
	u v |_{\Sigma_c} = -  \frac{4r_c}{\a}.
\end{equation}
The parameter $\epsilon$ represents the width of $\d(u v + 4 r_c / \a)$. In order to satisfy the Lorentz invariance condition, a particularly convenient candidate%
\footnote{There is an $\order{1}$ number uncertainty in choosing $\epsilon$; however, this number can be absorbed into a redefinition of the reduced Planck length $\tilde{\ell}_p$.  Also note that this $\epsilon$ is independent of the hydrodynamic expansion parameter $\varepsilon$.} %
for $\epsilon$ is that $\epsilon = 4 r_c / \alpha$. Setting $\epsilon = 4 r_c/\a$ gives us 
\begin{equation} \label{eq:delta inverse rc}
	\frac{4}{\a} \d \qty(u v + \frac{4 r_c}{\a}) \approx \frac{2}{\pi r_c}.
\end{equation}

Now we have all the ingredients to evaluate $\expval{H_{uu}^{2}}$. First, we apply the definition of the ``light ray'' operator~\eqref{eq:classical Xv} as in~\cite{Verlinde_2022,Zhang:2023mkf}, and perform a coordinate transformation to the ingoing Rindler coordinates $(\tau, r, x^i)$
\begin{align} \label{eq:Xv dot two point diffeo}
	\begin{split}
		\expval{H_{uu}^2} &\equiv \expval{\pd_u X^v(u, \bm{x}) \pd_{u'} X^v(u', \bm{x}')} \\
		&= \expval{\qty(\pdv{\tau}{u})\pd_{\tau} X^v (\tau, \bm{x}) \qty(\pdv{\tau'}{u'})\pd_{\tau'} X^v(\tau, \bm{x}')}\\
		&= \frac{4}{\a r_c} e^{-\a (\tau + \tau') / 2} \expval{\pd_{\tau} X^v (\tau, \bm{x}) \pd_{\tau'} X^v(\tau', \bm{x}')},
	\end{split}
\end{align}
where we have used $\pdv*{\tau}{u} = 2 / ( \a u)$ and the explicit relation between $u$ and $\tau$~\eqref{eq:u v diffeo}. 

In the next step, we substitute in the delta function of the ingoing Rindler time $\tau$~\eqref{eq:delta tau tau0}, as well as the regularized delta function of the radial coordinate $r$ in terms of a Poisson kernel of a width $\pi r_c/2$~\eqref{eq:delta inverse rc}
\begin{equation} \label{eq:XvXv langevin}
	\expval{\pd_{\tau} X^v (\tau, \bm{x}) \pd_{\tau'} X^v(\tau', \bm{x}')} = \frac{\ell_p^2 \a r_c}{8 \pi r_c } e^{\a (\tau + \tau') / 2} \delta(\tau - \tau')  f(\bm{x}; \bm{x}').
\end{equation}
Eq.~\eqref{eq:XvXv langevin} takes on the form of a first order Langevin equation. Although the factor of $r_c$ seems to cancel out between the chosen delta function regularization scheme and the coordinate transformation between $u$ and $\tau$, we will soon discover that another hidden factor of $r_c$ will show up once we relate the ingoing Rindler time $\tau$ back to the Minkowski global time $T$ relevant for a laboratory observer. 

In order to compute the two-point correlation function of $X^v$, we integrate both sides of~\eqref{eq:XvXv langevin} over $\tau'$ and $\tau$ readily to obtain 
\begin{align} \label{eq:two point Xv}
	\begin{split}
		\expval{X^v(\bm{x}) X^v(\bm{x}')} &= \frac{\ell_p^2 \a r_c}{8 \pi r_c} f(\bm{x}; \bm{x}')\int_{\tau_i}^{\tau_f} \dd{\tau} \int_{- \infty}^{\tau} \dd{\tau'} e^{\a(\tau + \tau') / 2} \d(\tau - \tau') \\
		&= \frac{\ell_p^2}{8 \pi} \qty(\frac{\a \D u^2}{r_c}) f(\bm{x}; \bm{x}'),
	\end{split}
\end{align}
where the Lorentz invariant combination $\a / r_c$ will be fixed in terms of the size of the causal diamond and the scale of quantum fluctuations shortly. The quantity 
\begin{equation} \label{eq:Delta u square}
	\D u^2 \equiv \frac{r_c}{\a} \qty(e^{\a \tau_f} - e^{\a \tau_i})
\end{equation}
denotes the square of the total elapsed coordinate time in $u$ according to the coordinate transformation~\eqref{eq:u v diffeo}. 
The light cone coordinate%
\footnote{Note that our convention of the light-cone coordinates differs from \cite{Zhang:2023mkf}. As a result, $(u,v)$ directions in Fig.~\ref{fig:cutoff} are flipped, compared to the corresponding figure in \cite{Zhang:2023mkf}.} %
$u$ is related to Minkowski coordinates $(T,Z)$ via $u = T + Z$. We immediately see that the elapsed time $\D u = \D T$ for an observer at a fixed location, \textit{e.g.} $Z =L$. Photon traversal time measured by a free falling observer on the boundary of a causal diamond of size $L$ is
\begin{equation} \label{eq:round trip time}
	\D T_{\rm r.t.} = 2 L = \D u
\end{equation} 
in the absence of any spacetime fluctuations.

Quantum fluctuations give rise to a ``smeared-out'' horizon, $\Sigma_c$, represented by the green curve Fig.~\ref{fig:cutoff}. This cutoff surface intersects with the boundary of the causal diamond at a distance $\tilde{\ell}_p$ from either bottom or top of the diamond. As in Refs.~\cite{Banks:2021jwj,Zurek:2022xzl,Verlinde_2022, Zhang:2023mkf}, $\tilde{\ell}_p$ characterizes the scale of quantum fluctuations. From Fig.~\ref{fig:cutoff}, it is clear that the Minkowski global time difference between the light front and $\Sigma_c$ is $\delta T = \tilde{\ell}_p$. This implies that $\d u = \tilde{\ell}_p$ for $Z = L$. Moreover, $\delta v = \delta u = \tilde{\ell}_p$ by symmetry. We choose to evaluate the uncertainty at the future tip of the diamond, at which future light ray intersects with the worldline of a free-falling observer at $Z = L$. These constraints along with Eq.~\eqref{eq:uv cutoff} imply that
\begin{equation} \label{eq:4rc/alpha}
		\frac{4 r_c}{\a}  = -(u_f - \d u) (v_f - \d v) \approx  u_f \d v 
		\approx 2 L \tilde{\ell}_p,
\end{equation}
where $(u_f , v_f)= ( 2 L, 0)$ corresponds to the location at the top of the diamond. Putting Eq.~\eqref{eq:4rc/alpha} back into Eq.~\eqref{eq:two point Xv} gives us the final answer 
\begin{equation} \label{eq:two point Xv x xprime}
	\expval{X^v(\bm{x}) X^v(\bm{x}')} = \frac{\ell_p^2}{\pi} \qty(\frac{L}{\tilde{\ell}_p}) f(\bm{x};\bm{x}').
\end{equation}
Since $X^v$ and $X^u$ are related via time-reversal symmetry $X^v \leftrightarrow X^u$, we immediately write down the two-point function of $\expval{X^u(\bm{x}) X^u(\bm{x}')}$ 
\begin{equation} \label{eq:two point Xu}
	\expval{X^u(\bm{x}) X^u (\bm{x}')} = \frac{\ell_p^2}{\pi} \qty(\frac{L}{\tilde{\ell}_p}) f(\bm{x};\bm{x}').
\end{equation}
The total time delay is the sum of Eqs.~\eqref{eq:two point Xv x xprime} and \eqref{eq:two point Xu}
\begin{align} \label{eq:variance time delay}
	\begin{split}
		\D T_{\rm r.t.}^2(\bm{x}; \bm{x}') &\equiv \expval{X^v(\bm{x}) X^v(\bm{x}')} + \expval{X^u(\bm{x}) X^u (\bm{x}')}  \\
		&= 2\frac{\ell_p^2}{ \pi} \qty(\frac{L}{\tilde{\ell}_p}) f(\bm{x}; \bm{x}').
	\end{split}
\end{align}
In four dimensions, there is no distinction between $\tilde{\ell}_p$ and $\ell_p$~\cite{Zurek:2022xzl, Verlinde_2022}; in other words, $\tilde{\ell}_p \cong \ell_p$, and Eq.~\eqref{eq:variance time delay} reduces to
\begin{equation}
	\Delta T_{\rm r.t.}^2(\bm{x}; \bm{x}') \stackrel{d = 4}{=} \frac{2\ell_p L}{\pi} f(\bm{x}; \bm{x}'). \label{eq:var time delay 4d}
\end{equation}
This equation is the main result of this section, and it is consistent with the findings from past literature~\cite{Verlinde:2019xfb,Verlinde:2019ade,Banks:2021jwj,Zurek:2022xzl,Verlinde_2022,Zhang:2023mkf} that used different and complementary means. Note that the variance in the photon traversal time~\eqref{eq:var time delay 4d} has an overall scaling dependence on both UV ($\ell_p$) and IR ($L$) scales. This kind of UV/IR mixing is most frequently encountered in the context of Brownian motion~\cite{feynman1964}. The hallmark of Brownian motion is that the variance of a Brownian particle depends on two quantities: 1) a diffusion coefficient $D$, an inter-molecular length scale proportional to the mean free path between collisions, and 2) total time elapsed $t$, usually set by laboratory measurement, takes on a macroscopic scale. In our case, we have found that quantum fluctuations near the light front of a Minkowski causal diamond also exhibit Brownian-like behavior. This result offers an exciting new avenue to test quantum fluctuations in gravity using precise laser interferometer measurement~\cite{Verlinde:2019ade,Bub:2023bfi}.

\section{Summary and Outlook}
\label{sec:summary}

We have studied the relation between fluid and shockwave geometries, showing via Petrov classification that a potential fluid contains the same physical information as a shockwave geometry.  We then proceeded to add a source to the fluid equation.  The stress energy source is quantum in nature, with an amplitude given by a fundamental uncertainty in spacetime.  We solved the equation of motion for spacetime and found an uncertainty in light travel time that depends both on the UV scale (a Planck length), as well as the light-crossing time of the causal diamond. 

We hope the relation between the shockwave and potential fluid dual geometries will allow for direct calculation in the sourced potential fluid itself, \emph{e.g.}~of the two point function for the potential fluid, including its dependence on time and transverse directions. Although technically more difficult, extending beyond the transverse planar limit should allow for understanding the angular dependence within finite causal diamonds. Ideally we could also use the fluctuation-dissipation theorem to compute the fluid diffusion constant from the quantum source.
 We leave these and other extensions to future work.

\acknowledgments
We thank Tom Banks for collaboration on related directions  \cite{Banks:2023wua}, as well as Lars Aalsma, Samarth Chawla, Tom Hartman, Temple He, Maulik Parikh, Jude Pereira, Allic Sivaramakrishnan, Marika Taylor, and Raphaela Wutte for insightful discussions. YZ and KZ are supported by the Heising-Simons Foundation “Observational Signatures of Quantum Gravity” collaboration grant 2021-2817, the U.S. Department
of Energy grant DE-SC0011632, and the Walter Burke Institute for Theoretical Physics.  KZ is also supported by a Simons Investigator award.  C.K. and S.-E.B. are supported by the U.S.
Department of Energy under grant number DE-SC0019470 and by the Heising-Simons Foundation
Observational Signatures of Quantum Gravity collaboration grant 2021-2818.

\appendix
\section{Derivation of the Brown-York Stress Tensor}
\label{app:BY Derivation}
In this appendix, we derive the Brown-York stress tensor in the non-relativistic hydrodynamic expansion~\eqref{eq:Tab explicit}.%
 To end up with this result, we first describe the near-equilibrium fluid metric by allowing the fluid variables $u_a, p$ to depend on $(\tau,x^i)$~\cite{Compere:2011dx,Compere:2012mt, Pinzani-Fokeeva:2014cka}. In the derivative expansion\footnote{The derivative expansion is defined as an order of different scaling $\tilde{\e}$ such that $\partial_r \sim \mathcal{O}(1),\,\partial_a \sim \mathcal{O}(\tilde{\e})$.}, the zeroth order seed metric is written as
\begin{align}\label{eq:zeroth oreder metric}
    d s^2=-2 p (x) u_a(x) d x^a d r+\left[\gamma_{a b}-\a p^2(x)\left(r-r_c\right) u_a (x) u_b(x)\right] d x^a d x^b
\end{align}
where $x$ denote $(\tau, x^i)$. 

In this work, we are interested in the non-relativistic hydrodynamic limit, which agrees with the near-horizon expansion~\cite{Bredberg:2011jq,Keeler:2020rcv}
\begin{equation} \label{eq:hydro scaling-Appendix}
	v_i \sim \order{\e}, \quad P \sim \order{\e^2}, \quad \pd_i \sim \order{\e}, \quad \pd_{\tau} \sim \order{\e^2}.
\end{equation}
In order to do that, consider the non-relativistic expansion of the pressure and four-velocity
\begin{align}\label{eq:from rel to nonrel}
    &p=\frac{1}{\sqrt{\alpha r_c-2P}}\approx\frac{1}{\sqrt{\alpha r_c}}+\frac{P}{(\alpha r_c)^{3 / 2}}+\mathcal{O}\left(\e^4\right)\,,\\
    & u_0=\frac{-\alpha r_c}{\sqrt{\alpha r_c-v^2}} \approx \sqrt{\alpha r_c}\left(1+\frac{v^2}{2 \alpha r_c}\right)+\mathcal{O}\left(\e^4\right)\,, \\
    & u_i=\frac{v_i}{\sqrt{\alpha r_c-v^2}} \approx \frac{v_i}{\sqrt{\alpha r_c}}+\mathcal{O}\left(\e^3\right)\,,
\end{align}
Here, $v_i$ is the non-relativistic fluid velocity in the transverse direction, and $v^2 \equiv v^i v_i$. $P$ is the non-relativistic pressure of the fluid on $\Sigma_c$.
We perform the non-relativistic hydrodynamic expansion in terms of the parameter $\e$, which keeps track of the scaling of various quantities. Combining these and expanding the metric in power of $\e$, the resulting metric becomes \cite{Bredberg:2011jq, Keeler:2020rcv}
\begin{align} \label{eq:hydro metric}
	\begin{split}
		d s^2 = &- (\a r) d \tau^2 + 2 d \tau d r + \delta_{i j} d x^i d x^j \\
		&- 2 \qty(1 - \frac{r}{r_c}) v_i dx^i d \tau - 2 \frac{v_i}{\a r_c} d x^i d r \\
		&+ \qty( 1 - \frac{r}{r_c} ) \qty[ (v^2 + 2 P) d \tau^2 + \frac{v_i v_j}{\a r_c} d x^i d x^j] + \qty(\frac{v^2}{\a r_c} + \frac{2 P}{\a r_c}) d \tau d r\\
            &-\frac{\left(r^2-r_c^2\right)}{\a r_c} \partial^2 v_i d x^i d \tau+\order{\e^3}.
	\end{split}
\end{align}
The unit normal vector on $\Sigma_c$ is given by~\cite{Bredberg:2011jq}
\begin{equation}\label{eq:unit normal}
	n^{\mu} \pd_{\mu} = \frac{1}{\sqrt{\a r_c}} \pd_{\tau} + \sqrt{\a r_c} \qty(1 - \frac{P}{\a r_c}) \pd_r + \frac{v^i}{\sqrt{\a r_c}} \pd_i + \order{\e^3},
\end{equation}
and after a short computation, we obtain the extrinsic curvature $K_{ab}= \frac{1}{2} \mathcal{L}_{n} \gamma_{a b}$ on $\Sigma_c$~\cite{Bredberg:2011jq}
\begin{equation} \label{eq:Kab}
    K_{ab}dx^a dx^b= -\frac{\a \sqrt{\a r_c}}{2} d\tau^2 +\frac{\a v_i}{\sqrt{\a r_c}}dx^i d\tau-\frac{\a(v^2+P)}{2\sqrt{\a r_c}}d\tau^2 -\frac{\a(v_i v_j -2 r_c \partial_{(i}v_{j)} )}{2(\a r_c)^{3/2}}dx^i dx^j+\mathcal{O}(\e^3)\,.
\end{equation}
Evaluating the Brown-York stress tensor defined in~\eqref{eq:Tab}, we obtain the Eq. \eqref{eq:Tab explicit}
\begin{equation} \label{eq:Tab explicit-Appendix}
    \kappa^2\, T_{a b} d x^a d x^b = \frac{\a v^2}{\sqrt{\a r_c}} d \tau^2  - \frac{2\a v_i}{\sqrt{\a r_c}} d x^i d \tau + \frac{\a}{\sqrt{\a r_c}} d x^i d x_i + \frac{P \delta_{i j} + v_i v_j }{r_c \sqrt{\a r_c}} d x^i d x^j - \frac{2 \pd_{(i} v_{j)} }{\sqrt{\a r_c}} d x^i d x^j + \order{\e^3}\,.
\end{equation}

\bibliographystyle{unsrt} 
\bibliography{References}

\end{document}